\documentclass[A4paper, 11 pt, conference]{ieeeconf}  

\IEEEoverridecommandlockouts       
\usepackage{geometry}
\usepackage{graphics} 
\usepackage{epsfig} 
\usepackage{mathptmx} 
\usepackage{times} 
\usepackage{amsmath} 
\usepackage{amssymb}  
\usepackage{subfigure}
\usepackage{algorithm} 
\usepackage{algorithmic} 
\usepackage{multirow} 
\usepackage{cite}
\usepackage{url}
\begin{document}

\title{\LARGE \bf
Cyber-Physical Authentication Scheme for Secure V2G Transactions}

\author{Yunwang Chen, Yanmin Zhao, and Siuming Yiu$^{*}$
\thanks{
$^*$Corresponding author: Siuming Yiu is with the Department of Computer Science, The University of Hong Kong, Hong Kong, China (email: smyiu@cs.hku.hk). 
}
\thanks{
Yunwang Chen is with the Department of Electrical and Electronic Engineering, Southern University of Science and Technology, Shenzhen, China (email: chenyw2021@mail.sustech.edu.cn).
}
\thanks{
Yanmin Zhao is with the Department of Computer Science, The University of Hong Kong, Hong Kong, China (emails: ymzhao@cs.hku.hk).
}
}

\maketitle
\thispagestyle{empty}
\begin{abstract}
The rapid global adoption of electric vehicles (EVs) demands robust cybersecurity measures for smart charging infrastructures. Despite protocols like ISO 15118 and OCPP utilizing public key infrastructure (PKI), trust issues persist due to fragmented certificate authorities and identity discrepancies. This paper proposes a blockchain-enabled authentication scheme to secure high-risk interactions like plug and charge (PnC). By integrating smart contracts between EVs, charging stations, and the PKI infrastructure within a localized consortium blockchain, the scheme enhances trust and security among stakeholders. A comprehensive security and privacy analysis demonstrates the protocol's effectiveness in mitigating risks such as man-in-the-middle (MitM) attacks and data breaches.
\end{abstract}


\section{Introduction}

The accelerating global adoption of electric vehicles (EVs) marks a significant transition towards sustainable transportation, driven by the pressing need to address the energy crisis, curb environmental pollution, and reduce reliance on fossil fuels. According to the International Energy Agency \cite{IEA2024}, global EV sales surged to nearly 14 million units in 2023, reflecting a 35\% increase from the previous year and bringing the total number of EVs on the road to over 40 million. This growth trajectory is expected to continue, with projected sales reaching 17 million in 2024, comprising over 20\% of all new vehicle sales worldwide. Moreover, the proliferation of EVs is intricately linked to the deployment of smart charging infrastructures that ensure stability, safety, and efficient energy management across electric grids \cite{Powell2022}.

However, the rapid expansion of EVs and their associated charging stations introduces new cyber-physical vulnerabilities, posing significant risks to both the power grid and the security of personal data. The integration of internet-connected EV charging station management systems (EVCSMS) expands the attack surface for cybercriminals, who may exploit system vulnerabilities to launch attacks that can destabilize the grid. For instance, malicious actors could exploit these vulnerabilities to orchestrate distributed denial of service (DDoS) attacks, disrupt charging processes, or compromise grid stability by manipulating EV charging patterns \cite{Sarieddine2023}. From a consumer standpoint, weaknesses in charging systems expose sensitive data such as payment details and account credentials, increasing the risk of financial fraud and identity theft \cite{Parameswarath2022}.

Recently disclosed cybersecurity vulnerabilities have demonstrated the growing risks associated with EV charging infrastructure. As revealed by the National Institute of Standards and Technology \cite{NVD23958}, a vulnerability in the Bluetooth communication system of certain EV charging devices allowed attackers to bypass authentication protocols, potentially granting unauthorized access to control charging parameters. Another instance involved a critical flaw in an EV charging software stack, which exposed charging infrastructure to unauthorized access, thereby increasing the risk of system manipulation \cite{NVD4622}. The potential for malicious actors to exploit these systems as they become more interconnected highlights the urgency of these security measures.

Given the cybersecurity risks associated with EV charging systems, the implementation of communication standards and protocols is crucial for protecting the stakeholders involved \cite{Gandhi2022}. Communication protocols or standards such as ISO 15118, Open Charge Point Protocol (OCPP), and GB/T 27930 are instrumental in ensuring secure and efficient interactions between EVs and charging stations \cite{Das2020}. These protocols optimize energy resource management and facilitate economically and energy-efficient charging processes while supporting the development of secure and user-friendly billing systems. To address evolving challenges and meet user demands, these standards have been updated with new features. For example, ISO 15118-2022 introduces the Plug and Charge (PnC) functionality, simplifying the charging process by enabling an EV to automatically authenticate, authorize charging sessions, and complete payments without manual intervention, much like the "charge and go" convenience of a credit card \cite{ISO2022}. OCPP 2.0 significantly enhances the security of electric vehicle charging infrastructure by building the public key infrastructure (PKI) for mutual authentication between EV chargers and management systems, utilizing digital certificates to prevent unauthorized access \cite{OCA2024}. Similarly, the GB/T 27930-2023 standard reflects China’s efforts to standardize and secure EV communication tailored to the specific needs of its rapidly growing market \cite{GB2023}.

While communication protocols such as ISO 15118 and OCPP, which utilize PKI, provide foundational security for EV charging, challenges remain in maintaining trust across the ecosystem \cite{Cecoin2020}. Given the vast market, no single certificate authority can monopolize trust, leading to fragmentation and inconsistencies. Additionally, discrepancies between real identities and users pose subtle but significant concerns. Moreover, delegating certificate issuance to subordinate agencies can lead to an imbalance of power, potentially compromising trust due to excessive authority granted to these entities. As a result, trust issues persist in interactions among EVs, charging stations, and the grid.

Recent advancements in distributed ledger systems, specifically permissioned blockchain, offer an additional layer of security and automation for energy transactions. In the context of EV charging, blockchain can work alongside existing PKI systems promoted by protocols like ISO 15118. With PKI providing the foundational security through digital certificates, a localized blockchain ledger can be deployed within a distributed power grid in a proximate region to ensure additional transparency \cite{Afzal2022}. This integration can help address the remaining trust issues between stakeholders by maintaining immutable records of transactions, thus providing a complementary mechanism to the existing PKI-based system. Moreover, blockchain, through smart contracts, can automate energy exchange, thereby securing the energy schedule as well as utility-based energy transactions \cite{Wan2022}.

In light of these challenges and opportunities, this paper aims to provide a holistic approach to secure EV integration with the grid by introducing a blockchain-enabled authentication scheme. Assuming a localized consortium blockchain is deployed within the distributed power grid, the proposed scheme focuses on protecting high-risk interactions like PnC while integrating smart contracts between EVs, charging stations, and the PKI. The main contributions of the paper are summarized as follows:
\begin{itemize}
    \item A cyber-physical PnC authentication method is developed to ensure tamper-proof and secure PnC interactions, safeguarding against unauthorized access and data manipulation.
    \item A smart contract is proposed to facilitate utility-based energy trading, with energy allocation calculated by the charging point based on utility functions from both the EV and the grid.
    \item A comprehensive security and privacy analysis is conducted, demonstrating the effectiveness of the protocol in mitigating risks such as man-in-the-middle (MitM) attacks within the distributed power grid.
\end{itemize}

The rest of the paper is organized as follows. Section II provides background information, including a review of V2G and blockchain applications in energy systems. Section III discusses the system model and threat model within a distributed power grid. Section IV presents the proposed authentication scheme and energy trading smart contract. Section V offers a detailed security and privacy analysis, and finally, Section VI concludes the paper.

\section{Background}

\subsection{Vehicle-to-Grid (V2G)}

The V2G concept involves how an EV communicates and interacts with the power grid. This communication can range from simple power signaling to more advanced information exchanges, such as transactions. When V2G is implemented, an EV can either supply energy to the grid or store energy, depending on whether the power flow is unidirectional or bidirectional in practical use. Unidirectional V2G allows for smart scheduling of EV charging, such as shifting demand to off-peak hours, thereby reducing grid pressure and charging costs. Meanwhile, bidirectional V2G represents an advanced paradigm where EVs function as distributed energy storage units. This enables EVs to help balance supply and demand, provide ancillary services like frequency regulation, load leveling, and voltage regulation, and generate profits for EV owners \cite{Azimi2021}. To support the V2G paradigm, various organizations have proposed different communication protocols, with ISO 15118, OCPP, and GB/T 27930 being among the most widely adopted protocols.

\subsection{Blockchain Applications in Energy Systems}

Blockchain technology, known for its security, transparency, and automation capabilities, has emerged as a promising solution for safeguarding information security among stakeholders in EV charging networks \cite{Noor2018}. As a distributed and immutable digital ledger, blockchain securely records transactions transparently and tamper-proof, ensuring data privacy, authenticity, and integrity within the network \cite{Yap2023}. 
The energy web decentralized operating system (EW-DOS) exemplifies how blockchain can be integrated with IoT to manage energy assets more effectively, offering more secure and reliable grid operations compared to conventional methods \cite{EnergyWeb2024}. 

Additionally, blockchain-based privacy-preserving schemes like V2GEx have been proposed to conduct secure and equitable electricity/service exchanges in V2G systems by utilizing zero-knowledge proofs and hash-chain micropayments to protect user identities and transaction details \cite{Wan2022}. Furthermore, blockchain facilitates secure and decentralized energy trading, enabling consumers to trade energy directly with each other and promoting distributed energy transactions as well as the operation of virtual grids \cite{Wu2023}. However, there remains a gap in blockchain-based protocols specifically designed for communication between EVs and EVCS, particularly those incorporating new features like PnC while supporting both scheduling and trading for energy transactions.

\section{Methodology}
\subsection{V2G System in Distributed Powergrid}

\begin{figure}[htbp]
\centerline{\includegraphics[width=3in]{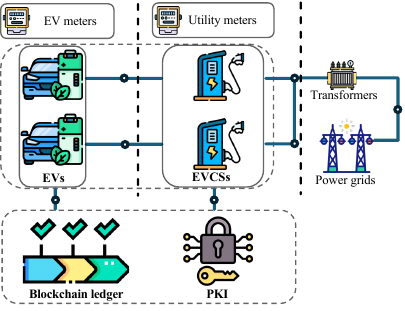}}
\caption{V2G system model.}
\label{fig:V2G_System_Model}
\end{figure}
\vspace{-10pt}
The V2G system model in this study is depicted in Fig. \ref{fig:V2G_System_Model} and aligns with the ISO 15118 standard for entities and processes involved in Plug-and-Charge (PnC) authentication for EV charging. The key entities in this model include EVs equipped with an electric vehicle communication controller (EVCC), EV charging stations (EVCSs) with a supply equipment communication controller (SECC), and various backend systems supporting PnC operations.

\textbf{EVs:} In V2G systems, EVs act as prosumers, entities that produce and consume energy. With storage capabilities, EVs can inject energy back into the grid or draw energy as needed. The decision-making process of EVs is influenced by economic incentives, adjusting their energy dispatch in response to dynamic pricing and demand signals. Each EV is inherently equipped with metering capabilities to track energy consumption and injection, and uses the EVCC to communicate with the SECC at the EVCS via ISO 15118. The EVCC establishes a PnC session, manages the EV’s credentials, and serves as the endpoint for EV authentication.

\textbf{EVCS:} The SECC at the EVCS facilitates data transfer between the EVCC and the backend systems built by the utility company. It authorizes the vehicle to use the charging service based on credentials issued by the utility company. The SECC plays a crucial role in ensuring secure communication and managing the overall charging process, enabling seamless PnC sessions between the EV and the backend systems.

\textbf{Blockchain network:} The EVCS performs scheduling and trading through smart contracts on a consortium blockchain network with connected EVs being its clients. This blockchain network is deployed in the distributed power grid within a proximate region, ensuring that energy transactions, including those related to charging and discharging activities, are executed transparently and fast.

\textbf{PKI:} The PKI framework utilizes digital certificates to ensure the authenticity and integrity of the information exchanged within the PnC sessions. This infrastructure is supported by certificate authorities (CAs), responsible for issuing, distributing, revoking, and installing these certificates across EVs, EVCSs, and utility companies. CAs also encompass certificate provisioning services (CPS), which validate and provision these credentials, ensuring that only authenticated entities can participate in the network.

\textbf{Utility company:} In the V2G ecosystem, the utility company functions as an aggregator responsible for managing EVCSs, mediating between individual EV owners and the power grid. It aggregates the energy capacities of a fleet of EVs to buy or sell energy in bulk, enhancing flexibility and reliability in microgrids. The utility company also manages contracts, redistributes payments to EV owners, and secures profits through V2G administration fees.

\subsection{Threat Model}

The ISO 15118 standard establishes security measures to protect the PnC process through secure communication over TLS channels and the use of strong cryptography. We assume that, under normal conditions, the communication cannot be compromised, and the cryptographic protections remain intact unless an attacker gains access to the necessary keys. This threat model considers two primary types of attacker: those with full physical access to the EVCS and those with remote access. Each of these methods can be compromised to undermine the PnC process.

\textbf{Malicious EVCS:} A compromised EVCS could broadcast fraudulent charging parameters, such as fake availability or reduced charging power, to deceive EVs and their owners. It could modify charging parameters after reservations are made, potentially leading to overcharging, undercharging, or fraudulent billing. Additionally, a malicious EVCS could attempt to extract, copy, or duplicate PnC credentials from connected EVs, including contract certificates, OEM provisioning certificates, and other private keys.

\textbf{Malicious EV:} A compromised EV might agree to charge at an EVCS but subsequently refuse to pay, causing financial losses and operational disruptions for the charging station. It may also steal credentials from other legitimate vehicles, use them for unauthorized charging, or falsely report charging/discharging parameters, potentially destabilizing the grid. The EV could attempt to relay authorization requests to the EVCS to illicitly obtain signed credentials or manipulate charging sessions for its benefit.

\subsection{Communication Settings}

The EVs and the EVCSs communicate through two types of channels. Wireless channels are used for transmission of information between the EV, EVCS, and PKI, which verifies the EV’s and EV’s identity. A physical CAN-bus link is a wired connection that transmits physical challenges, ensuring that the EV is physically connected to the EVCS and facilitating the metering data exchange.

\section{Proposed Authentication Scheme}
The proposed cyber-physical authentication scheme within the V2G system serves to verify both the digital identity of the EV and its physical connection to the designated EVCS. This dual verification mechanism is essential to preventing unauthorized access and ensuring the integrity of the PnC process, as outlined by the ISO 15118 standard. The process is divided into three main phases: {Registration}, {Authentication}, and {Transaction}.

\subsection{Registration}

The user registers with the PKI by submitting a verified identifier $ID_{{user}}$ (e.g., a government-issued ID) and generating a pseudo-identifier $PID_{{user}}$ with a digital wallet for ongoing transactions. 

The PKI issues verifiable credentials $VC_{{user}}$ and $VC_{{EVCS}}$ to legitimate users and EVCSs, respectively, which are immutably stored on the blockchain network. To maintain user privacy, the PKI generates a series of pseudo-IDs $PID_{{user}}^i$ mapped to the original pseudo-ID $PID_{{user}}$. Each $PID_{{user}}^i$ is unique to a single V2G session to prevent correlation by potential adversaries. The PKI securely stores these $PID_{{user}}^i$ and their associated cryptographic key pairs $K_{{user}}^i = (K_{{user}}^{i,{public}}, K_{{user}}^{i,{private}})$ for future verification. As for legitimate EVCSs, the PKI will issue EVCS ID $ID_{{EVCS}}$ along with cryptographic key pairs $K_{{EVCS}} = (K_{{EVCS}}^{{public}}, K_{{EVCS}}^{{private}})$ and verifiable credential $VC_{{EVCS}}$. Note that $ID_{{EVCS}}$, $K_{{EVCS}}$, and $VC_{{EVCS}}$ are updated regularly.
\cite{Parameswarath2022}.

\subsection{Authentication}

\begin{figure}[htbp]
\centerline{\includegraphics[width=3in]{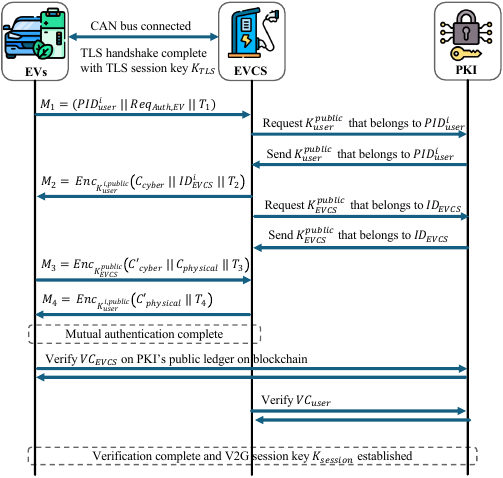}}
\caption{Authentication process.}
\label{fig:Authentication_Process}
\end{figure}

When the EV is plugged into the EVCS, a TLS handshake will be performed, during which a TLS session key $K_{{TLS}}$ will be derived. Before each V2G session, both the EV and the EVCS must authenticate each other through the following steps, as illustrated in Fig. \ref{fig:Authentication_Process}:

In step 1, the EV initiates the authentication by selecting a pseudo-ID $PID_{{user}}^i$ and the corresponding cryptographic key pair $(K_{{user}}^{i,{public}}, K_{{user}}^{i,{private}})$ for the session. The EV sends $PID_{{user}}^i$, a request $Req_{{(Auth,EV)}}$ for authentication, and a timestamp $T_1$ to the EVCS, forming $M_1 = (PID_{{user}}^i || Req_{{(Auth,EV)}} || T_1)$.

In step 2, the EVCS retrieves the user’s public key $K_{{user}}^{i,{public}}$ from the PKI and generates a cryptographic challenge $C_{{cyber}}$. This challenge $C_{{cyber}}$, along with EVCS ID $ID_{{EVCS}}$ and a hashed timestamp $T_2 = H(T_1)$, where $H$ is a cryptographic hash function agreed upon by the EVCS and EV, are encrypted with the user’s public key $K_{{user}}^{i,{public}}$ and sent back to the user as $M_2 = {Enc}_{K_{{user}}^{i,{public}}}(C_{{cyber}} || ID_{{EVCS}} || T_2)$.

In step 3, the EV decrypts the cyber challenge $C'_{{cyber}}$, EVCS ID $ID'_{{EVCS}}$, and $T'_2$ from ${Dec}_{K_{{user}}^{i,{private}}}(M_2)$. If $T'_2 = H(T_1)$, the EV communicates with the PKI to retrieve EVCS’s public key $K_{{EVCS}}^{{public}}$ based on $ID'_{{EVCS}}$. The EV responds to the EVCS with $C'_{{cyber}}$, a challenge $C_{{physical}}$ for the EVCS, and a timestamp $T_3 = H(T_2)$, encrypted with $K_{{EVCS}}^{{public}}$, as $M_3 = {Enc}_{K_{{EVCS}}^{{public}}}(C'_{{cyber}} || C_{{physical}} || T_3)$.

In step 4, the EVCS retrieves $C''_{{cyber}}$, $C'_{{physical}}$, and $T'_3$ from ${Dec}_{K_{{EVCS}}^{{private}}}(M_3)$. If $C''_{{cyber}} = C_{{cyber}}$ and $T'_3 = H(T_2)$, the EVCS responds to the EV with $C'_{{physical}}$ and a timestamp $T_4 = H(T_3)$ as $M_4 = {Enc}_{K_{{user}}^{i,{public}}}(C'_{{physical}} || T_4)$. When the EV recovers $C''_{{physical}}$ and $T'_4$ from ${Dec}_{K_{{user}}^{i,{private}}}(M_4)$, if $C''_{{physical}} = C_{{physical}}$ and $T'_4 = H(T_3)$, the mutual authentication is complete.

In step 5, the EV verifies the EVCS’s credentials by checking its verifiable credential $VC_{{EVCS}}$ against the blockchain public ledger. The EVCS verifies the user’s credentials by checking the pseudo-ID $PID_{{user}}^i$ and the signed credential $VC_{{user}}$ presented by the user against the set of hash values $H = h(PID_{{user}}^i)$ retrieved from the PKI. Once identities and credentials are verified, both the user and the EVCS establish a secure V2G session key $K_{{session}} = {Enc}_{K_{{user}}^{{private}}}(C_{{cyber}} \parallel C_{{physical}})$ from the combined challenges $C_{{cyber}}$ and $C_{{physical}}$ to encrypt subsequent data during the transaction phase.
\vspace{-4pt}
\vspace{-12pt}
\subsection{Transaction}

During the V2G process, when the EV \( v_i \) is plugged into the EVCS at time slot \( t \), a smart contract is automatically triggered to manage the transaction. The transaction revolves around the utility functions of both the EV and the EVCS, and follows a series of predefined steps:

\begin{enumerate}
    \item \textit{Utility function submission:} The EV \( v_i \) submits its utility function \( U_{{user},i}^{t}(x_i^{t}, p_c^t, p_d^t) \) to the EVCS through the smart contract. Here, \( x_i^{t} \) represents the amount of electricity to be charged or discharged from time slot \( t \) to \( t+1 \), and \( p_c^t \) and \( p_d^t \) represent the charging/discharging prices, respectively. The EVCS also has its own utility function \( U_{{EVCS},i}^{t}(p_c^t, p_d^t) \), which it uses to schedule and optimize the energy exchange for the period.
    
    \item \textit{Computation and scheduling:} The smart contract takes the submitted utility function \( U_{{EV},i}^{t} \) and combines it with \( U_{{EVCS},i}^{t} \) to compute the optimal energy exchange \( x_i^{t} \) that maximizes the joint utility. This calculation is done according to predefined rules in the smart contract, ensuring that the solution adheres to both the EV’s and EVCS’s constraints.
    
    \item \textit{Energy transfer monitoring:} During time slot \( t \) to \( t+1 \), the smart contract oversees the continuous energy transfer monitoring. Both the EV and EVCS report their metering data \( E_{{meter},{EV}} \) and \( E_{{meter},{EVCS}} \) to the smart contract. The smart contract checks the reported values to ensure that the discrepancy \( |E_{{meter},{EV}} - E_{{meter},{EVCS}}| \) is within an acceptable tolerance \( \Delta E_{{meter}} \).
    
    \item \textit{Billing and verification:} At time slot \( t+1 \), the smart contract generates a bill for the energy exchanged between the EV \( v_i \) and EVCS during the period from \( t \) to \( t+1 \), as well as other service fees.
    
    \item \textit{Transaction finalization:} If both the EV and EVCS agree on the billing amount, the smart contract automatically finalizes the transaction. The final transaction details, including the amount of energy transferred, the billing amount, and the time of completion, are recorded on the blockchain. This ensures that the transaction is transparent, immutable, and non-repudiable.
    
    \item \textit{Dispute resolution:} If there is a discrepancy outside the allowed tolerance \( \Delta E_{{meter}} \) or the EV and EVCS are unable to agree on the billing amount, the smart contract initiates a dispute resolution process as defined in its code. The contract may trigger further verification steps or refer to additional data inputs before reaching a resolution.
    
    \item \textit{Logging and next cycle preparation:} Upon completion of the transaction, the smart contract logs the details on the blockchain and prepares for the next operational cycle. The EV will also update its internal ledger with the latest transaction information, which will be ready for the subsequent cycle.
\end{enumerate}

\section{Security and Privacy Analysis}

In this section, we analyze the security properties of the proposed V2G system, focusing on how it addresses the threats outlined in our threat model. The analysis covers various attack scenarios, including potential vulnerabilities in the authentication protocol, smart contract operations, and energy transfer processes.

\subsection{Security Analysis}

\textbf{Public ledger attack:} The use of a blockchain network to log transactions between EVs and EVCSs introduces the possibility of public ledger attacks. An adversary might attempt to track transactions by analyzing patterns in public addresses and the amounts transferred. However, our system mitigates this threat by utilizing pseudo-identifiers \( PID_{user} \) that are unique for each transaction and unlinked to any previous or future transactions. This prevents meaningful correlation between transactions and specific users. Additionally, the EVCS credentials will be updated regularly.

\textbf{MitM attack:} In a MitM attack, an adversary could attempt to intercept communications between the EV and EVCS during the authentication phase. Our system mitigates this by employing end-to-end encryption using keys derived during the initial registration phase. Specifically, during each session, the EV and EVCS exchange cryptographic challenges \( C_{{physical}} \) and \( C_{{cyber}} \), which are encrypted using the public keys of the communicating parties. The integrity of the communication is ensured by verifying these challenges with the private keys held securely by each party. This encryption, coupled with session-specific pseudo-identifiers, ensures that even if an adversary intercepts the communication, they cannot decrypt or tamper with the data without the necessary private keys.

\textbf{Replay attack:} Replay attacks involve the reuse of valid data transmissions by an adversary to impersonate an EV or EVCS. Our system prevents such attacks by including unique, time-sensitive cryptographic challenges during each authentication session. These challenges incorporate timestamps and are validated within the smart contract to ensure they are fresh and valid only within a specific time window. If an adversary attempts to reuse an old challenge, the smart contract will recognize it as expired and reject the transaction. Additionally, the session keys derived from these challenges are used to encrypt subsequent communications, ensuring that even if old transmission data is intercepted, it cannot be reused in a new session.

\textbf{Charging station or EV attack:} A compromised EVCS or EV could potentially manipulate charging parameters, extract credentials, or execute unauthorized transactions. Our system mitigates these risks by enforcing mutual authentication between the EV and EVCS before any transaction can proceed. During the authentication process, both the EV and EVCS must verify their credentials by successfully completing a challenge-response mechanism using their private keys. The smart contract that governs the transaction is designed to execute only after successful mutual authentication and metering within the allowed tolerance, thereby preventing unauthorized transactions or manipulation of billing amounts or the energy transferred, further mitigating the risk of fraudulent activities that could destabilize the grid.

\vspace{-10pt}
\subsection{Privacy Analysis}

Maintaining user privacy is a core component of the proposed V2G system. The use of pseudo-identifiers for each transaction ensures that the EV’s identity remains anonymous throughout the process. Additionally, the blockchain records transactions in a manner that does not reveal the true identity of the EV, further protecting user privacy. The following points highlight the privacy-preserving aspects of the system:

\textbf{EVCS Privacy:} The EVCS is only aware of the pseudo-ID associated with the transaction and cannot link this to the actual identity of the EV.

\textbf{Transaction Anonymity:} Each transaction is associated with a pseudo-ID and standard transaction amount, preventing correlation between multiple transactions.

\textbf{Data Minimization:} Only the necessary data for the transaction is shared between the EV and EVCS, with all sensitive information, such as the actual identity of the EV, remaining encrypted and inaccessible to unauthorized parties.

\section{Conclusion}
This paper has presented a cyber-physical authentication scheme for securing plug and charge (PnC) operations at the distributed power grid level. By combining cryptographic techniques with cyber-physical verification and blockchain-based smart contracts, the proposed protocol ensures secure, transparent, and efficient energy transactions between electric vehicles and charging stations, complementing the public key infrastructure. The comprehensive security and privacy analyses demonstrate the system's effectiveness against various threats, including distributed denial of service (DDoS), man-in-the-middle (MitM), and replay attacks. The proposed protocol provides a solution to the emerging cybersecurity challenges in EV charging systems, facilitating the secure integration of EVs into the power grid while preserving user privacy and ensuring reliable energy transactions.
\section*{Acknowledgment}
This work was supported by the
Undergraduate Training Programs for Innovation and Entrepreneurship of
Southern University of Science and Technology under Project 2024S14.

\bibliographystyle{IEEEtran}

\bibliography{references}

\end{document}